# *Deep Learning for Quantitative Dynamic Fragmentation Analysis*

[1]Erwin Cazares, [1*]Brian E Schuster

[1]Department of Metallurgical, Materials and Biomedical Engineering at the University of Texas at El Paso

**Abstract**

We have developed an image-based convolutional neural network (CNN) that is applicable for quantitative time-resolved measurements of the fragmentation behavior of opaque brittle materials using ultra-high speed optical imaging. This model extends previous work on the U-net model, where we trained binary, 3 and 5 class models using supervised learning on experimentally measured dynamic fracture experiments on various opaque structural ceramic materials that were adhered on transparent polymer (polycarbonate or acrylic) backing materials. Full details of the experimental investigations are outside the scope of this manuscript but briefly, several different ceramics were loaded using spatially and time-varying mechanical loads to induce inelastic deformation and fracture processes that were recorded at frequencies as high as 5 MHz using high speed optical imaging. These experiments provided a rich and diverse dataset that includes many of the common fracture modes found in static and dynamic fracture including cone cracking, median cracking, comminution, and combined complex failure modes that involve effectively simultaneous activation and propagation of multiple fragmentation modes. While the training data presented here was obtained from dynamic fragmentation experiments, this study is applicable to static loading of these materials as the crack speeds typically higher a kilometer per second in these materials are on the order of 1-10 km/s regardless of the loading rate. We believe the methodologies presented here will be useful in quantifying the failure processes in structural materials for protection applications and can be used for direct validation of engineering models used in design.

**Introduction**

*Corresponding author: bschuster@utep.edu

In the late 1960's, Wilkins and coworkers published a series of reports[1–3] on the dynamic properties of lightweight armor materials using an early application of ultrafast diagnostics including flash X-ray radiography and high-speed imaging that provided the first time-resolved measurements of the crater formation and fracture modes of metallic and brittle materials. Their systematic investigations combining experiments and computational modeling sought to identify the threshold stresses and stress states that controlled the response of these materials under complex loading states. In the spirit of this early work, the goal of the present work is to provide a toolset for quantitative real-time measurements of the distribution of fragment sizes under dynamic loading.

A clear understanding of the inelastic deformation and failure processes of these materials are required for the parameterization of phenomenological strength and damage models used in the design of materials and systems for extreme mechanical loading conditions. It is well documented in the literature that ceramic materials including SiC, B4C, alumina and others show controlling strengths that are highly pressure dependent. As summarized by Holmquist and Johnson[4], SiC-B fails at a Von Mises equivalent stress of 0.77 GPa during spall[5] while the strengths approach 13 GPa on high pressure and shock loading paths. Under complex loading experienced in transient loading conditions that can include severe spatially and time varying stresses that can caused fragmentation. There is an inherent need for ultra-fast characterization under both dynamic and static loading as the fracture velocities in these materials can greatly vary from in these materials are on the order of 1-10 km/s regardless of the loading rate[6,7].

Let us consider some of the characteristic fracture modes that can be activated during transient mechanical loading of brittles materials. For a moment will consider a highly idealized physical model for the high velocity impact of a blunt/spherical metallic body into a finite thickness brittle material system. Figure 1 summarizes four different fracture modes that can be activated during impact, crater formation and fragmentation of brittle material systems. At the lowest velocities and peak stresses (see Figure 1A), Hertzian cone cracks[8–10] have been found to initiate at the edge of the area of contact at the impact surface that propagate towards the back surface of the target. High velocity impact generates high intensity spherical waves into the target materials; depending on the details of the boundary conditions, these waves can reflect from free surfaces and interfaces in the materials system that result in a stress state that is nominally in equibiaxial. Related experiments in the literature indicate that equibiaxial stress can initiate cracks at the back surface that propagate back to the impact surface[11–16] that show similarities to Figure 1B. At higher stresses from impact, it is possible to generate median or penny shaped cracks that propagate away from the impactor to target interface (Figure 1C). Hypervelocity impact and impacts at the highest stresses can generate a comminuted region that propagates away from the impact site (Figure 1D).

Above we considered some possible failure modes, however, the real failure processes in brittle materials are much more complicated[1,17,18]. Typically quantitative measurements of the fragmentation behavior have focused on highly detailed post-mortem static measurements[12,13]. Material strengths depend on the loading path (or Von Mises equivalent stress) and stochastic failure processes[19,20] are related to the statistical distribution of defects in each individual sample batch[11]. Fragment sizes can be depend on the distance from the impact site where a fine comminuted zone closest to the zone transitions to coarser fragments further from the point of impact. Finally, target and impactor geometry can alter the time-history of stress and failure. The training data examined in this manuscript shows mixed modes of failure and we made no real attempt to clearly quantify the difference between these different failure modes. We focused instead on the rigorous assessment of binary, 3 and 5 classes semantic segmentation models applicable for quantitative analysis of high-speed optical imaging data from fracture experiments on brittle materials systems.

*Computer-vision-based and image segmentation approaches for crack detection*

The structural health monitoring community has been active in the development and application of algorithms for quantitative measurements of cracks to assess reliability in civil infrastructures. Early monitoring applications include those based on ruled-based algorithms[21–23], edge-detection algorithms[23,24], image thresholding[23,25,26] and morphology operations which are more difficult to adapt to the real-world applications due to the variable appearance of crack features and environmental conditions[23]. Such approaches required constant fine tuning of the systems that limits its resilience to new data and thus limits its deployment. Therefore, the use of machine learning has been highly researched for this specific task of identifying and classifying cracks on concrete infrastructure.

The goal for a digital imaging process is to extract substantial crack features that allows the computer to comprehend and analyze the given scene[23]. These processes include the image acquisitions, and image pre-processing to enhance or optimize contrast and/or brightness. Computer vision algorithms can use thresholding, erosion, dilation and other techniques to classify the cracks[27]. An example of this was done by Prasanna et Al.[28] where they introduced a spatially tuned robust multi-feature classifier (STRUM) that utilizes a series of feature computations by intensity-based and gradient-based features that are later used as input in a random-forest model to make crack classifications.

Deep learning applications are data-driven methods that do not require a manually designed pipeline of image operations, and instead the process of building a model summarizes to selecting a suitable network structure (a series of layers, connections, and mappings), a function to evaluate the model performance (loss function) and a reasonable optimization algorithm. More recently, deep learning methods have been applied in civil engineering to measure and quantify road damage[28–30], fracture at surfaces[31–33] and structure damage[34–36] to automate aspects of prioritizing service and maintenance. To our knowledge similar models have not been applied to dynamic or time-series fracture events with the goal of analyzing fragment sizes.

There are two main approaches for the detection of concrete cracks using deep learning technology. The first approach is based on the method of object detection which is a computer vision task where all the objects of interest in the image are located using a bounding box (rectangular frame) to determine its position[31,37]. However, this approach has limited recognition accuracy results when surface cracks have a high aspect ratio, varying orientation, thinner features scales, brightness variations, and gross differences in the surface-to-crack ratio. In practice, one can have difficulty with object detection in large bounding boxes where the crack appears as small fraction of the image area (e.g. less than 2%). Approaches that use object detection leverage the models to determine whether a crack exists within the image and results in classification problems. The second method using semantic segmentation distinguishes different types of defined classes by classifying each pixel in the image[38]. Here, our classes consist of cracks, bulk material, and background. With pixel level classification, we perform subsequent measurements on the predicted (class-assigned) mask for each defined class.

Crack detection systems must be able to recognize and locate crack morphologies within a given background. To achieve this deep learning semantic segmentation approaches are often used to classify pixel-to-pixel in the provided images. We selected this approach to our problem since most cracks are present as thin, angled features which are difficult to threshold based on surface ratio to the background of the samples. In the literature, this is achieved using deep learning approaches that apply convolutional layers and encoder-decoder models. Other common practices use transfer learning to help with training times. The model architecture we based our project is based on the U-Net approach, since it has shown remarkable results within literature using small and highly imbalanced datasets that resemble our application. The extensively varying real-world situations such as lighting and shadow changes bring challenges for crack detection. We built ensemble models to take advantage of the different backbones that generate unique features maps and yield another set of predictions for comparison.

**Methodology**

*Deep learning convolutional neural network architectures*

The network architectures used in this manuscript are based on U-Net segmentation model developed by Ronnenberg and coworkers[39]. This network has proven to have excellent results with few training samples, as it leverages the skip connections at the encoding phase for spatial feature retention. The original model consists of a series of convolutional, and max pooling layers that reduce the image size by half at each block. Then a series of convolutional transpose layers are used for reconstructing the image and taking input from the skip connections at the encoder phase. The results are produced from the last convolutional layer where a sigmoid or SoftMax function is utilized (depending on the number of classes) to predict a segmented which has the same size as the input.

We altered the original U-Net to include a batch normalization layer that is introduced between the convolutional blocks and the activation layers. This is done because it reduces internal covariate shift by normalizing the activations of each layer[40]. This prevents activations from becoming too large or too small, which can lead to vanishing or exploding gradients during training. We added a dense layer to our model at the bridge between the encoder and decoder phase to introduce non-linearity to the model, enabling it to learn more complex relationships between features. The addition of the dense layer increased the capacity of the model, allowing it to learn more discriminative representations of the input data. Subsequently, other approaches from well-known architectures were introduced using a Python library called *Segmentation Models*[41]. These networks are available to use in a U-Net configuration and includes the following backbones: VGG16, VGG19, ResNet (18, 34 and 101), SeResNet (50 and 152), and lastly DenseNet (101 and 121). We assessed the predictions for each architecture and used combinations later in an ensemble model.

*Data preparation and preprocessing techniques*

The original dataset contains a series of 47 different experiments where a high-speed camera was used to capture 128 or 256 images taken at interframe spacing larger than 200 nanoseconds. We did not specifically consider any details of the event time and associated framing rate. 300 images were selected from this dataset to build the training dataset of the model where images were selected based on the resolution, fragment geometry, and ease of crack visibility. Training focused on images captured early in the experimental event time, as it was commonly found that gross deformation following massive fragmentation causes strong gradients in contrast and brightness that do not allow fragment resolution. A flat field correction was applied to individual images to use all of the available gray scale. To reduce the memory and process larger batches, we changed the images from resolution from 16 to 8-bit encoding. The cracks were manually annotated using an iPad Pro model 2021. Figure 3A and B (respectively) presents the post-processed experimental image and the associated annotated binary mask.

At later times in most experiments, we observed that the interface between the ceramic target and transparent backing material would fail and separate at the outer edge of the circular target. The delaminated region had a pronounced increase in brightness and the associated cracks were not recognized by the binary models. Figure 3C shows the mask for the 3 class model consisting of the background, inner area cracks, and outer area cracks. Finally, Figure 3D shows the 5 class model that includes inner and outer fragments as additional labeled classes. In application of this model to analyze the materials response, it would be logical to debate if the radial cracks are in subsequently fragmented into inner and outer fragments, however, this will require more in-depth analysis that is not considered here.

*Experimental settings*

Models were run in an Anaconda Python environment on a Dell G15 laptop equipped with an Intel 10870H CPU with 16 GB of RAM memory and an RTX 3060 laptop GPU. All models were trained using an input shape of 256 x 416 to satisfy all model tensor shapes. Pixels were added as background to the original 250 x 400 pixel images and were removed after prediction. Models were run for 60 epochs in 4-image batches with an Adam optimizer using a learning rate of 0.0001 and 0.001. We used a custom loss function that will be defined below. Model training data was saved after training and subsequent metrics were calculated using a 100-image validation dataset. Metrics obtained from the latter were saved to build the ensemble models described in another section. Neither data augmentation techniques nor callbacks were utilized. The architectures selected were trained following the schema: models were trained as a set of two. Using the as described batches, learning rates and custom loss function. First, our two modified models (as described in the previous section) were trained and afterwards a set of backbones were adapted in the U-Net format from the segmentation models library.

*Evaluation criteria*

We utilized a training custom loss (Eq. 1) which is a combination of binary cross-entropy (BCE) and intersection over union (IoU) metrics, using the Keras package. BCE focuses on pixel-wise accuracy, penalizing the model for misclassifying individual pixels. This is important for precise delineation of object boundaries. While IoU, on the other hand, measures the overlap between predicted and true positives (TP) regions, providing region-wise accuracy, ensuring that the general structure of segmented regions is accurate. Combining the two encourages the model to correctly classify the pixels and capture the overall morphology of the segmented regions.

We used the accuracy, precision, and recall (Eq. 2-4, respectively) as the metrics to evaluate the performance of a model. Accuracy provides a general measure of how well the model is performing across all classes. However, it might not be the most informative metric for imbalanced datasets, in the context of our task, only about 6-10% of the pixels belong to the labeled classes (binary and 3-Class datasets). Therefore, the use of precision and recall is crucial in our segmentation tasks where false positives (FP) and false negatives (FN) (incorrectly predicted pixels) can have significant consequences. For example, increasing the crack thickness results in reduced fragment areas. Vice versa, reducing the crack thickness results in agglomeration of fragments.

Similarly, IoU, F1 Score and Dice coefficients (Eq. 5-7) were also used as model metrics. The F1 Score is the harmonic mean of precision and recall and offers a single metric to ensures that neither false positives nor false negatives disproportionately influence the evaluation. This is particularly critical in our scenario were missing a crack (FN) can be as detrimental as incorrectly identifying a non-crack region as a crack (FP). The Dice coefficient, closely related to the F1 Score, measures the similarity between the predicted and ground truth segments. It is especially useful in the context of image segmentation where the spatial overlap of the predicted and actual regions is the prediction goal. By incorporating the Dice coefficient, we ensure that our model is not only accurate in terms of individual pixel classifications but also in the spatial coherence of the segmented regions. This comprehensive evaluation approach was done for developing robust segmentation models capable of accurately identifying and delineating cracks especially in our imbalanced dataset.

(1) $\text{CustomLoss} = \text{Binary Cross} - \text{Entropy} + (1 - \text{Intersection Over Union})$

(2) $\text{Accuracy} = (\text{\# of correct preds})/(\text{total \# of preds})$

(3) $\text{Precision} = \text{True Positives} / (\text{True Positives} + \text{False Positives})$

(4) $\text{Recall} = \text{True Positives} / (\text{True Positives} + \text{False Negatives})$

(5) $F1 = (Precision + Recall)/(2 \times Precision \times Recall)$

(6) $\text{Dice} = (2\ x\ (\text{True Positives} + \text{False Positives} + \text{False Negatives})) / (2 \times \text{True Positives})$

(7) $\text{IoU} = (\text{True Positives}) / ((\text{True Positives} + \text{False Positives} + \text{False Negatives}))$

*Model assessment and ensemble building*

Model performance was assessed using a 100-image validation dataset. Where all previously defined metrics were calculated. These metrics were saved for each model and tracked with the prediction type for binary and multiclass classification. From such metrics the seven best performing models on recall and IoU (which are positively correlated) chosen to build the ensemble models.

Three different ensemble model strategies were applied in this study. The first strategy is a hard voting ensemble, where the mean of model predictions determines the class. A threshold value of 0.5 is applied, resulting in a majority vote of the class as the predicted label. The second strategy utilizes a soft voting method, which averages the class probabilities. Using the last model layer, a threshold value of 0.5 is applied to the mean class probabilities, voting on the label before a sigmoidal or SoftMax function is applied. Lastly, a Bayesian model averaging (BMA) technique was used, weighting averages towards more confident models based on recall and IoU metrics. These weights were determined experimentally as 0.25, 0.25, 0.20, 0.10, 0.10, 0.05, and 0.05, respectively, for building the binary, 3-class, and 5-class models.

Subsequently, a more robust ensemble model was created from each hard ensemble prediction type. Based on testing, only the BMA approach was used to build the high-end model. The weights for this model were 0.6, 0.3, and 0.1, respectively. Preliminary results indicated that the binary model is well-suited for detecting inner cracks with high resolution, the 3-class model effectively adds the detection of outer cracks seen at later stages, and the 5-class model introduces fragment recognition. The robust ensemble model was then selected for subsequent fragment analysis due to its superior recall accuracy as seen in Table 1.

Once the predictions are made with the ensemble models, we use the connected components functions from OpenCV Python[42] to calculate and measure the distinct fragments. We consider a fragment as the areas enclosed by the predicted cracks. We calculate unique features based on pixels that include the area, perimeter, aspect ratio, roundness, orientation (respect to the sample's center) and bounding box coordinates. A pseudo color scheme is introduced to return a color image for all fragments detected. This data is collected and saved under a csv file for future analysis.

**Results and Discussion**

*Model validation metrics*

We assessed the ability of model to accurately discern and differentiate complex structures, such as overlapping cracks or irregular fragments, with the metrics outlined in the previous section. Table 1 provides a comprehensive overview of the metrics obtained by the highest-performing model and the three distinct ensemble strategies. Our primary objective in constructing the ensemble model was to enhance both recall and mean IoU metrics. Notably, while ensemble voting for individual tasks yielded only marginal improvements, the robust ensemble, defined as the all-ensemble model, produced a significant 5% increase in the recall metric, reaching a total of 94.54%. This substantial enhancement underscores the efficacy of the ensemble approach in augmenting model performance as more delineating crack pixels can be introduced for later fragment size calculations.

We are using different segmentation tasks to promote a more refined understanding of the scene by introducing more distinct labeled classes. This refinement was expected to enhance the model's ability to delineate challenging structures within the sample's surface. However, we observed a trade-off in precision when analyzing larger area aspects such as fragments, due to the increased classes that classifies sample surface into fragments.

The consistency of IoU scores (seen in Table1) across different segmentation tasks is a notable observation in our analysis. This stability can be attributed to several factors: the models demonstrate robust generalization capabilities, effectively applying learned patterns and features across diverse segmentation tasks and that the features extracted by the models appear to be consistently relevant across different tasks, contributing to consistent performance in IoU calculations. This suggests that the architecture and training procedures effectively prioritize and utilize features essential for crack segmentation, resulting in stable performance metrics.

*Model predictions*

It is important to highlight that despite constraints posed by the problem statement and limitations such as image resolution, focus, and lighting, our results exceed expectations, were the goal was to segment the cracks and perform subsequent fragment size analysis. Overall, we have seen that all models are capable of segmenting cracks with variable recall and IoU scores; these models can predict in all instances of the whole dataset within reasonable limits. Figure 4 illustrates a comparison between predicted outcomes. Across all models, inner cracks predictions remain present and consistent, while multiclass segmentation enhances the clarity of fragment edges, sample surface and adds the outbounded area present at later stages. Consequently, all predicted cracks were converted to a binary mask, as pinpointing fragment location and classification falls beyond the scope of this project.

In Figure 4A, the input image provided to the model depicts a sample with mixed mode fracture with features consistent with cone cracking superimposed with radial cracking initiated either from the impactor to target interface and/or the back surface of the target (see Figure 1A-C). This image was partially out of focus on the right side, a common observation in this series and other experiments. Figure 4B displays the input mask generated by the 3-Class model, accurately labeling distinguishable cracks. Figure 4C shows the results of the binary ensemble model where the predicted cracks closely reflect the geometry of the inner fragments from the mask with distinctions made solely on open crack segments. Figure 4D expands the prediction to include the outer area, capturing some partial cracks. Figure 4E offers the addition of the inner and outer fragment area, while Figure 4F presents the ensemble of all models, eliminating the differentiation between outer and inner areas.

When running the model across the entire dataset, we observe successful prediction in each scene, facilitating the implementation of fragment analysis. However, there are two exceptional scenarios where the models encounter difficulties. The first occurs at the initial stages of each series when crack formation is underway. During this phase, cracks do not intersect and therefore the connected components algorithm is not able to define the enclosed area that defines a fragment (as seen in Figure 5A). In this case, the model tracks the entire sample surface as a single unique "fragment." An example of this complete time series is included in the supplementary materials. The second scenario arises when the sample is highly comminuted, typically observed towards the end of a series. Here, the fragments are too small (less than ~10 pixels) for the model to detect accurately. Such instances of underperformance were excluded from metrics calculations. It is in view of the authors that in later applications of this model, users can alter the optical imaging conditions and field of view to study fragments sizes that control the response for their experimental conditions of interest.

*Fragment analysis on different response mechanisms*

We assessed the performance of our robust model across various material responses depicted in the dataset, as illustrated in Figure 5. Each scenario showcased highlights specific challenges and capabilities of our model in studying brittle material responses. In Figure 5B, the model is challenged to predict thin and long radial fragments with additional rotation. This example illustrates the strength of semantic segmentation over object recognition to accurately capture the full extent of the cracks, including their shape, size, and orientation, which are essential for comprehensive analysis and accurate prediction of material behavior. Figure 5C presents a case study featuring a predominant cone and secondary cone cracking. Here, the central features exhibit complex geometries of many sizes, forming the characteristic cone around the center while thicker radial fragments emerge from the cone, intersected by a secondary cone response, introducing challenges for accurate crack delineation. Figure 5D shows a common response within the dataset, where the material exhibits a mixed mode between radial and secondary cone cracking. This scenario demonstrates the model's capability to adapt to varying crack patterns and modes of failure that arise in these sub-processes. Finally, Figure 5E depicts the later stage of most experiments, where a highly comminuted area is observed. Here, the model begins to agglomerate the detected fragments and as material comminution progresses there is an increasing fraction of unresolvable areas from a combination of variation in lighting conditions across scenes and the resolution limitations in discerning smaller fragment sizes. It is important to emphasize that while our model serves as a valuable tool for fragment analysis through image segmentation, its scope is limited to segmentation tasks only. Subsequent labeling of these fragments is necessary to achieve comprehensive classification of material's response.

*Time-series analysis*

We conducted a comprehensive case study utilizing a complete time series to analyze fragment sizes, with the aim of evaluating the model's practical deployment. Leveraging the predictions of our robust model and subsequent postprocessing techniques outlined in our methodology, we obtained and stored fragment data at each frame index. Figure 6 illustrates the analysis conducted for an experiment showcasing mixed-mode failure, encompassing cone, secondary cone, and radial cracking, culminating in a comminuted central area and outbound characteristics. Figures 6A-C depict the input at distinct stages of the sample, while Figures 6D-E display the corresponding fragment prediction masks.

The distribution of fragment size in all scenarios demonstrates a log-normal histogram that indicates that primary and subsequent fractures typically result in a higher number of smaller fragments as the material breaks down over time. The accompanying graph demonstrates a decreasing trend in mean fragment size alongside a rise in fragment count, while maintaining relative consistency in the studied area (sample surface area) compared to the original scene. This relationship validates the predictions from the model from the expected experimental response, as there is a clear time dependence to the fragment size where primary and subsequent fracture is expected to reduce the fragment size as a function of time.

In later stages of the experiment, the mean fragment size approaches the global minimum which highly depends on the stress time history observed by the target. The inset within the graph reveals that despite the decrease in fragment size, variability persists between subsequent frames. It is noted that the smallest fragment size detected is ~10 pixels, as this threshold is employed to minimize mask noise during fragment calculation. Also, a drop in fragment count and predicted area is observed as the outbounded area becomes unresolvable, and challenges arise to predict fragments while increasingly number of finer comminuted grains fall below the detectable size. While Figure 6 includes a plot of the time history of the average fragment size, the supplementary materials include the associated histograms of the fragment size and aspect ratios. Post-processing algorithms can be applied to the segmented connected components fragments to track a large list of possible fragment attributes that are not considered further here.

This analysis section highlights the importance of model performance for deployment. It underlines the need for the model to consistently predict the evolution of fragment sizes to guarantee precise comprehension and analysis of material behavior. This reinforces the potential of the model to understand material behavior while ensuring accurate prediction of the corresponding sample area. Furthermore, it emphasizes the scalability of similar analyses for individual experiments, which can be stored for subsequent examination and analysis. We highlight the limitations of the model in instances where gross deformation and flow of the failed materials and fragments may lead to missing or agglomerated predictions. These scenarios arise from the lack of direct control over imaging conditions and the timing of the experiments. Despite these challenges, our goal is to provide a robust model capable of offering a quantitative tool for fragment segmentation.

**Conclusions**

In this project, we have introduced a novel application of deep learning to conduct fragment size analysis, aimed at studying brittle material response under transient and dynamic mechanical loading. Our investigation encompasses various segmentation tasks, including binary, 3-class, and 5-class models, leading up to a robust ensemble model that achieves a 5% increase in recall accuracy. Leveraging the U-Net framework, our model has demonstrated capabilities in segmenting and analyzing fragments, particularly in scenarios involving cone, secondary cone, and radial cracking, with a minimum fragment area of 10 pixels. We present innovative techniques in image segmentation and postprocessing, enabling us to examine diverse material behaviors and crack patterns observed in impact experimental datasets captured through high-speed imaging. Our analyses highlight the critical importance of model performance in accurately predicting the evolution of fragments where the early stages of loading show a sharp gradient in fragment sizes. These findings underscore the necessity for consistent and reliable predictions to accurately understand and analyze material behavior. Our detailed case study illustrates a new ability to quantify the evolution in fragment count and size during transient mechanical loading that offers valuable insight into the time-dependent failure processes for materials.

**Supplementary materials.**

The supplementary materials include results on an experimental series showcasing the fragments and 2D histogram plots. These plots illustrate the relationship between fragment size and fragment aspect ratio, providing a detailed visual analysis of the fragment characteristics and evolution of the segmented fragments.

**Acknowledgments**

This work was supported by the U.S. Army Combat Capabilities Development Command Army Research Laboratory (DEVCOM ARL) under Cooperative Agreement Number W911NF2020194. The authors would like to acknowledge Dr. Phillip Jannotti, Dr. Christopher Hoppel, Dr. Lionel Vargas-Gonzalez and Dr. Brandon McWilliams.

**Conflict of interest statement**

The authors have no conflicts to disclose.

**Data availability statements**

The data supporting the training of the models and their predictions masks used in this study are subject to third-party restrictions imposed by DEVCOM ARL Laboratory. Access to the data requires permission from the respective third party.

**Author Contributions**

**Erwin Cazares:** Methodology (lead); Software (lead); Formal analysis (lead); Validation (lead); Visualization (lead); writing – original draft (lead). **Brian Schuster:** Conceptualization (lead); Funding acquisitions (lead); Resources (lead); Supervision (lead); Project Administration (lead); Review, and editing (lead).

# References

1. Wilkins ML, Cline CF, Honodel CA. *FOURTH PROGRESS REPORT OF LIGHT ARMOR PROGRAM*. California Univ., Livermore. Lawrence Radiation Lab.; 1969. doi:10.2172/4173151

2. Wilkins M. *Second Progress Report of Light Armor Program*. Lawrence Livermore National Lab. (LLNL), Livermore, CA (United States); 1967. doi:10.2172/7156835

3. Third Progress Report of Light Armor Program. Accessed May 11, 2024. https://apps.dtic.mil/sti/citations/tr/AD1033497

4. Holmquist TJ, Johnson GR. Characterization and evaluation of silicon carbide for high-velocity impact. *J Appl Phys*. 2005;97(9):093502. doi:10.1063/1.1881798

5. D. P. Dandekar, P. T. Bartkowski. Report No. ARL-TR-2430. Published online March 2001.

6. Ji M, Li H, Zheng J, Yang S, Zaiemyekeh Z, Hogan JD. An experimental study on the strain-rate-dependent compressive and tensile response of an alumina ceramic. *Ceram Int*. 2022;48(19, Part A):28121-28134. doi:10.1016/j.ceramint.2022.06.117

7. Lamberson L, Eliasson V, Rosakis AJ. In Situ Optical Investigations of Hypervelocity Impact Induced Dynamic Fracture. *Exp Mech*. 2012;52(2):161-170. doi:10.1007/s11340-011-9521-0

8. Chaudhri MM, Liangyi C. The orientation of the Hertzian cone crack in soda-lime glass formed by oblique dynamic and quasi-static loading with a hard sphere. *J Mater Sci*. 1989;24(10):3441-3448. doi:10.1007/BF02385722

9. Frank FC, Lawn BR. On the theory of Hertzian fracture. *Proc R Soc Lond Ser Math Phys Sci*. 1997;299(1458):291-306. doi:10.1098/rspa.1967.0137

10. Kocer C, Collins RE. Angle of Hertzian Cone Cracks. *J Am Ceram Soc*. 1998;81(7):1736-1742. doi:10.1111/j.1151-2916.1998.tb02542.x

11. Collini L, Carfagni GR. Flexural strength of glass–ceramic for structural applications. *J Eur Ceram Soc*. 2014;34(11):2675-2685. doi:10.1016/j.jeurceramsoc.2013.10.032

12. Tomba-Martinez AG, Cavalieri AL. Fracture Analyses of Alumina Subjected to Mechanical and Thermal Shock Biaxial Stresses. *J Am Ceram Soc*. 2002;85(4):921-926. doi:10.1111/j.1151-2916.2002.tb00193.x

13. Nie X, Chen WW, Templeton DW. Dynamic Ring-on-Ring Equibiaxial Flexural Strength of Borosilicate Glass. *Int J Appl Ceram Technol*. 2010;7(5):616-624. doi:10.1111/j.1744-7402.2010.02508.x

14. Swab JJ, Shoulders WT. Equibiaxial flexure strength and fractography of long-wave infrared materials. *J Eur Ceram Soc*. 2023;43(15):7059-7067. doi:10.1016/j.jeurceramsoc.2023.07.023

15. Swab JJ, Patel PJ, Tran X, et al. Equibiaxial Flexure Strength of Glass: Influence of Glass Plate Size and Equibiaxial Ring Ratio. *Int J Appl Glass Sci*. 2014;5(4):384-392. doi:10.1111/ijag.12094

16. Yıldız BK, Tür YK. An investigation of equibiaxial flexural strength and hardness properties of Al2O3–Ni nanocomposites based microstructures with ZrO2 and Cr2O3 additives. *Mater Sci Eng A*. 2019;758:103-111. doi:10.1016/j.msea.2019.05.014

17. Sathananthan P, Sirois A, Singh D, Cronin D. Sphere on tile ballistic impact experiment to characterize the response of soda lime glass. Int J Impact Eng. 2019;133:103321. doi:10.1016/j.ijimpeng.2019.103321

18. Shockey DA, Marchand AH, Skaggs SR, Cort GE, Burkett MW, Parker R. Failure phenomenology of confined ceramic targets and impacting rods. *Int J Impact Eng*. 1990;9(3):263-275. doi:10.1016/0734-743X(90)90002-D

19. Zavattieri PD, Espinosa HD. Grain level analysis of crack initiation and propagation in brittle materials. *Acta Mater*. 2001;49(20):4291-4311. doi:10.1016/S1359-6454(01)00292-0

20. Espinosa HD, Zavattieri PD. A grain level model for the study of failure initiation and evolution in polycrystalline brittle materials. Part II: Numerical examples. *Mech Mater*. 2003;35(3):365-394. doi:10.1016/S0167-6636(02)00287-9

21. Kittler J, Illingworth J, Föglein J. Threshold selection based on a simple image statistic. *Comput Vis Graph Image Process*. 1985;30(2):125-147. doi:10.1016/0734-189X(85)90093-3

22. Koutsopoulos HN, Downey AB. Primitive-Based Classification of Pavement Cracking Images. *J Transp Eng*. 1993;119(3):402-418. doi:10.1061/(ASCE)0733-947X(1993)119:3(402)

23. Kheradmandi N, Mehranfar V. A critical review and comparative study on image segmentation-based techniques for pavement crack detection. *Constr Build Mater*. 2022;321:126162. doi:10.1016/j.conbuildmat.2021.126162

24. Guo F, Qian Y, Liu J, Yu H. Pavement crack detection based on transformer network. *Autom Constr*. 2023;145:104646. doi:10.1016/j.autcon.2022.104646

25. (PDF) Image-Based Crack Detection Using Sub-image Technique. Accessed May 11, 2024. https://www.researchgate.net/publication/339266767_Image-Based_Crack_Detection_Using_Sub-image_Technique

26. Unified Approach to Pavement Crack and Sealed Crack Detection Using Preclassification Based on Transfer Learning | Journal of Computing in Civil Engineering | Vol 32, No 2. Accessed May 11, 2024. https://ascelibrary.org/doi/full/10.1061/%28ASCE%29CP.1943-5487.0000736

27. Wang P, Chen P, Yuan Y, et al. Understanding Convolution for Semantic Segmentation. In: *2018 IEEE Winter Conference on Applications of Computer Vision (WACV)*. ; 2018:1451-1460. doi:10.1109/WACV.2018.00163

28. Prasanna P, Dana KJ, Gucunski N, et al. Automated Crack Detection on Concrete Bridges. *IEEE Trans Autom Sci Eng*. 2016;13(2):591-599. doi:10.1109/TASE.2014.2354314

29. Ji H, Kim J, Hwang S, Park E. Automated Crack Detection via Semantic Segmentation Approaches Using Advanced U-Net Architecture. *Intell Autom Soft Comput*. 2022;34(1):593-607. doi:10.32604/iasc.2022.024405

30. Ji A, Xue X, Wang Y, Luo X, Xue W. An integrated approach to automatic pixel-level crack detection and quantification of asphalt pavement. *Autom Constr*. 2020;114:103176. doi:10.1016/j.autcon.2020.103176

31. Yamane T, Chun P jo. Crack Detection from a Concrete Surface Image Based on Semantic Segmentation Using Deep Learning. *J Adv Concr Technol*. 2020;18(9):493-504. doi:10.3151/jact.18.493

32. Wang W, Su C. Semi-supervised semantic segmentation network for surface crack detection. *Autom Constr*. 2021;128:103786. doi:10.1016/j.autcon.2021.103786

33. Junior GS, Ferreira J, Millán-Arias C, Daniel R, Junior AC. Ceramic Cracks Segmentation with Deep Learning. *Appl Sci*. 2021;11(13):6017. doi:10.3390/app11136017

34. Islam MMM, Kim JM. Vision-Based Autonomous Crack Detection of Concrete Structures Using a Fully Convolutional Encoder–Decoder Network. *Sensors*. 2019;19(19):4251. doi:10.3390/s19194251

35. Liu Z, Cao Y, Wang Y, Wang W. Computer vision-based concrete crack detection using U-net fully convolutional networks. *Autom Constr*. 2019;104:129-139. doi:10.1016/j.autcon.2019.04.005

36. Dung CV, Anh LD. Autonomous concrete crack detection using deep fully convolutional neural network. *Autom Constr*. 2019;99:52-58. doi:10.1016/j.autcon.2018.11.028

37. Usamentiaga R, Lema DG, Pedrayes OD, Garcia DF. Automated Surface Defect Detection in Metals: A Comparative Review of Object Detection and Semantic Segmentation Using Deep Learning. *IEEE Trans Ind Appl*. 2022;58(3):4203-4213. doi:10.1109/TIA.2022.3151560

38. Panella F, Lipani A, Boehm J. Semantic segmentation of cracks: Data challenges and architecture. *Autom Constr*. 2022;135:104110. doi:10.1016/j.autcon.2021.104110

39. Ronneberger O, Fischer P, Brox T. U-Net: Convolutional Networks for Biomedical Image Segmentation. In: Navab N, Hornegger J, Wells WM, Frangi AF, eds. *Medical Image Computing and Computer-Assisted Intervention – MICCAI 2015*. Lecture Notes in Computer Science. Springer International Publishing; 2015:234-241. doi:10.1007/978-3-319-24574-4_28

40. Ioffe S, Szegedy C. Batch normalization: accelerating deep network training by reducing internal covariate shift. In: *Proceedings of the 32nd International Conference on International Conference on Machine Learning - Volume 37*. ICML'15. JMLR.org; 2015:448-456.

41. Iakubovskii P. qubvel/segmentation_models. Published online May 10, 2024. Accessed May 11, 2024. https://github.com/qubvel/segmentation_models

42. Bradski G. The OpenCV Library. *Dr Dobbs J Softw Tools*. Published online 2000.

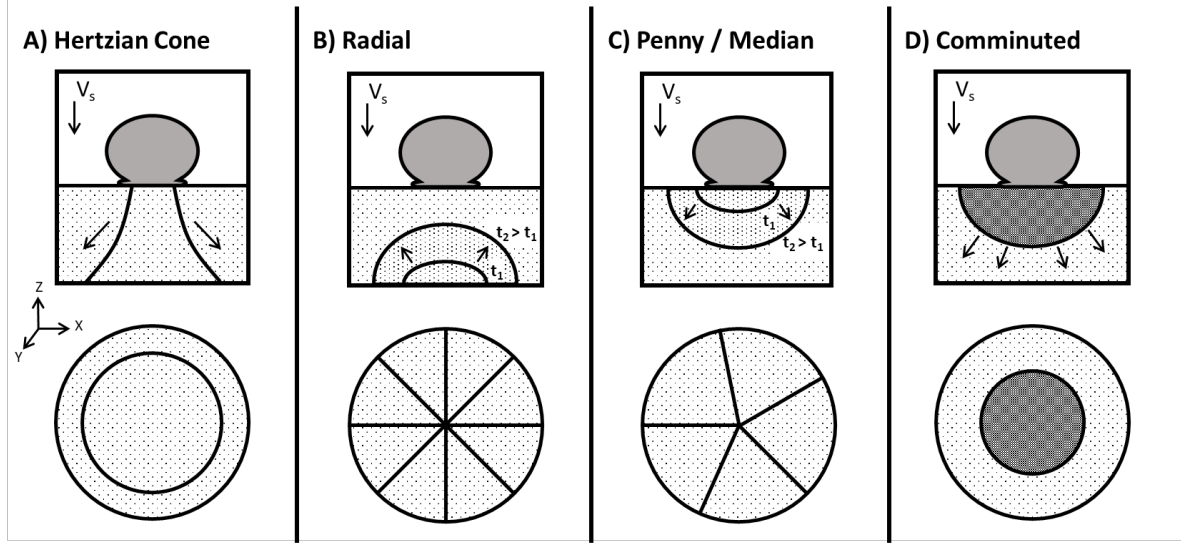

**Figure 1.** Schematic of select fracture modes found during transient mechanical and impact loading of brittle materials. The majority of experiments examined here display combined modes of failure from 2 or more of these failure modes. A) Under low velocity impact Hertzian cone cracks initiate at the impact surface and propagate in the impact direction. B) Reflection of high intensity spherical waves from the back surface of the target can initiate radial cracks at the back surface of the sample from tension and biaxial tension C) Higher stresses result in median or penny-shaped cracks that initiate from the impact surface. D) Under hypervelocity impact, a comminuted region forms and spreads from the impact site.

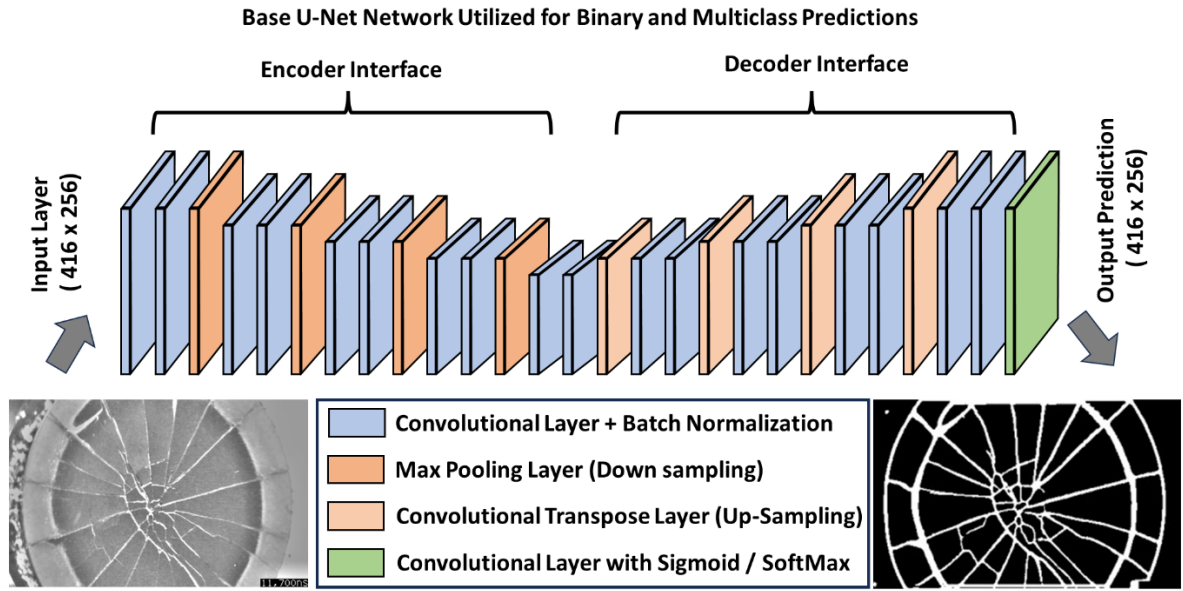

**Figure 2.** The modified convolutional model used here follows the U-Net structure of a contracting path (encoder phase) with a series of convolutional, batch normalization and max pooling layers followed by an extending path (decoder phase) where the inverse is done to produce a segmentation mask. The model is adapted from Ronnenberg et al architecture.

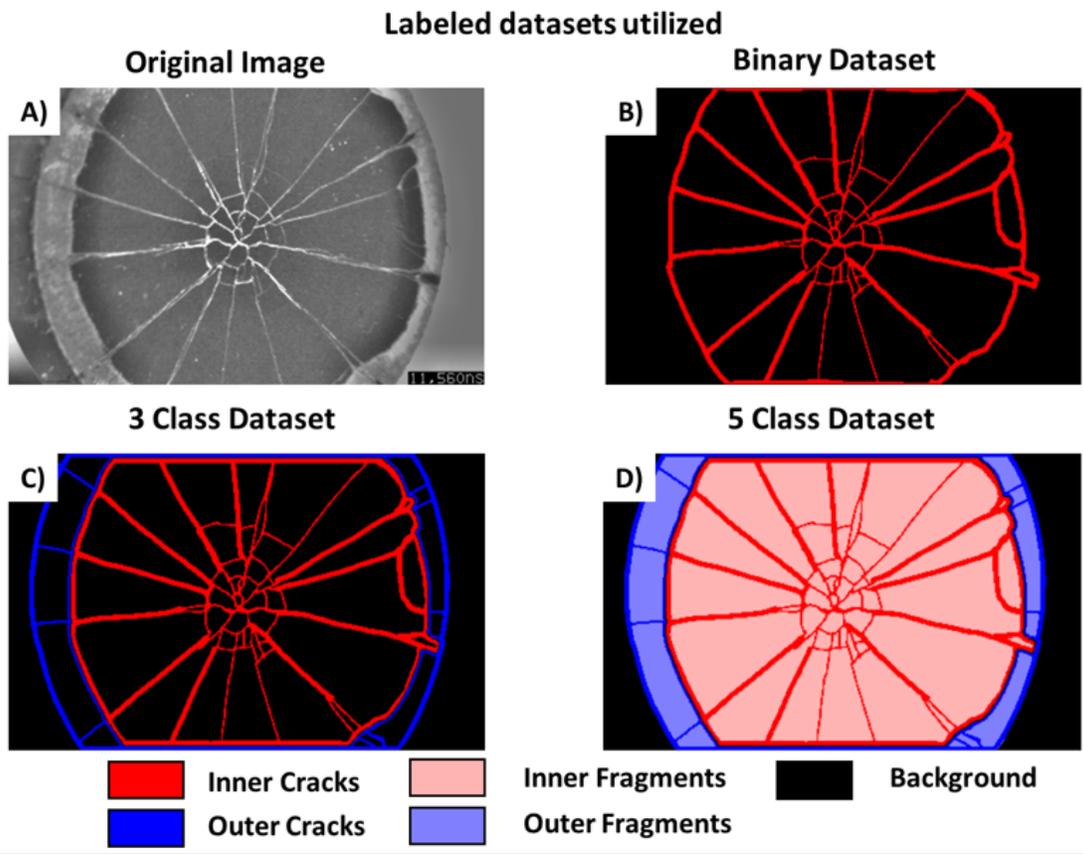

**Figure 3.** Different labeled datasets utilized for the training of the models. Where the original images were hand-labeled using an iPad Pro 2021 and annotation software. Section B shows the binary masks representing the cracks available, while C and D represent the 3 and 5 class datasets, respectively. The latter were added to aid in the segmentation tasks at a later time, a distinction is made from inner and outer areas.

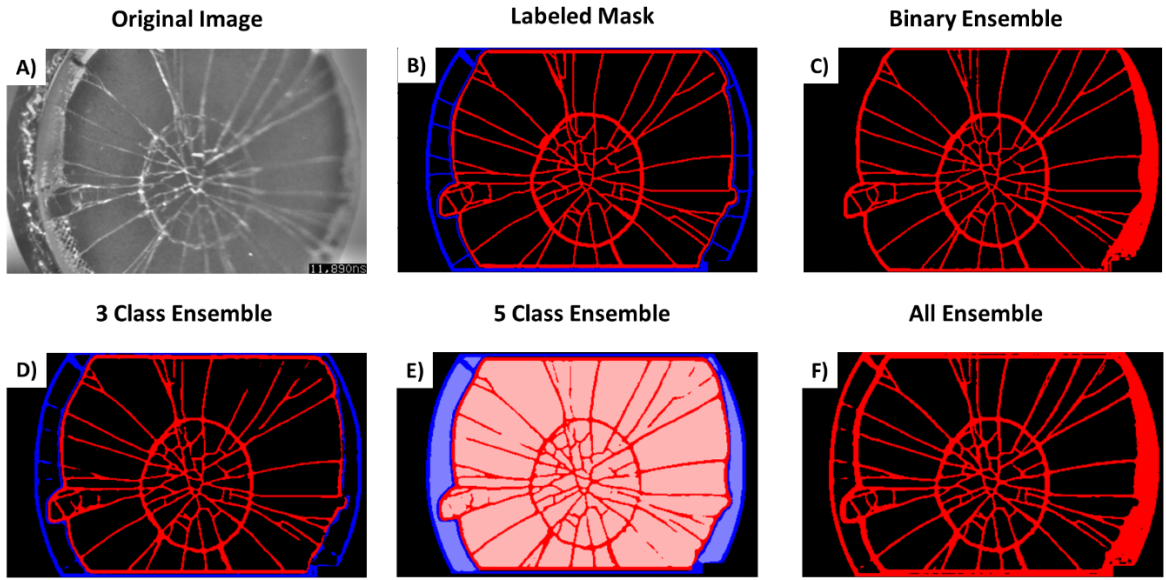

**Figure 4.** Model predictions utilizing the BMA ensemble models for various segmentation tasks. A and B depict the original input provided to the model, while C corresponds to the binary prediction focusing solely on inner cracks. D and E display the multiclass segmentation, incorporating labels for the outbound areas (outer cracks/fragments), and F illustrates the robust ensemble model formed from the ensemble models.

## Prediction on Different Fracture Modes

### A) Early Time Cracking
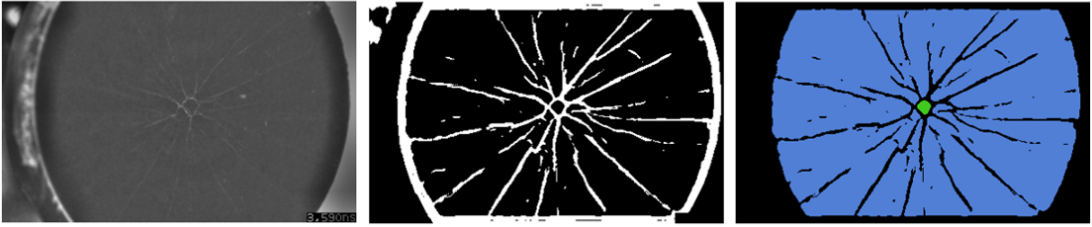

### B) Radial Cracking
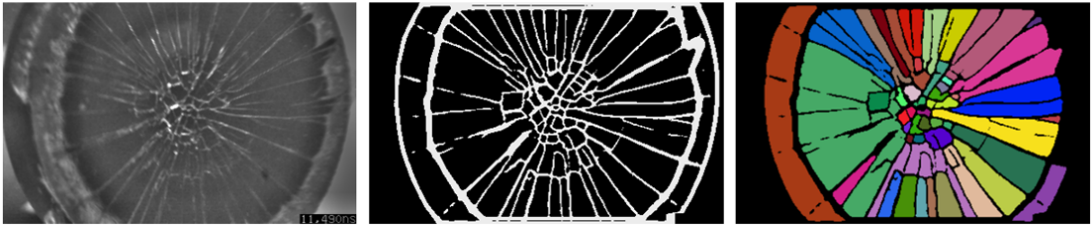

### C) Cone and Radial Crackings
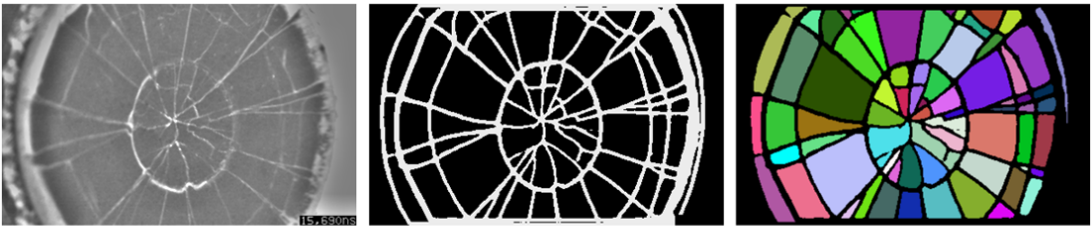

### D) Mixed Mode Cracking
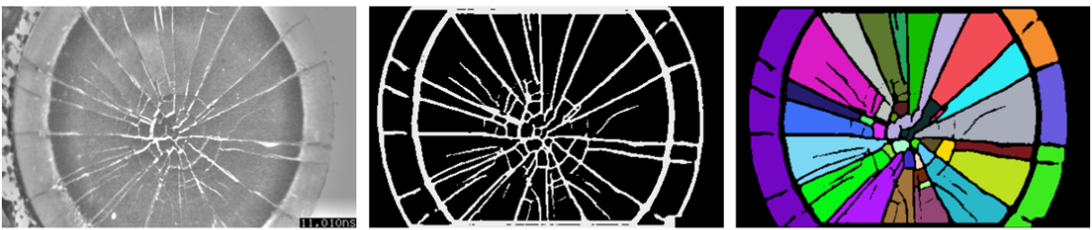

### E) Comminution
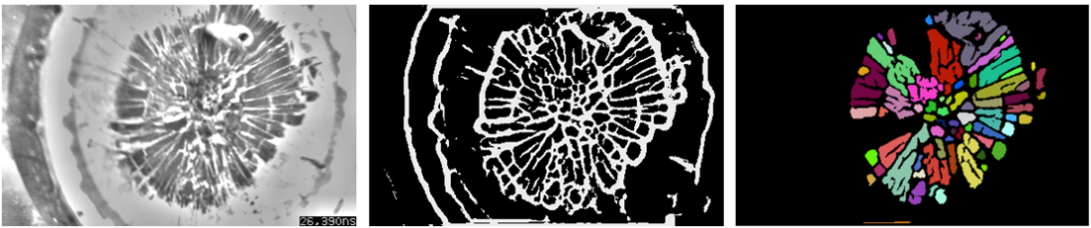

**Figure 5.** Evaluation of model predictions across various material responses, showcasing its capability to detect fragments at different experiment timeframes and mechanical behaviors. (A) At an early time, crack had initiated however clear connected component fragments were not detected. (B) Thin and long radial fragments with additional rotation. (C) Introduction of cone cracking and intersection of the radial components by a secondary cone crack. (D) Common response exhibiting mixed mode between cone and radial cracking. (E) Later stage showing highly comminuted area, where fragment predictions become increasingly agglomerated.

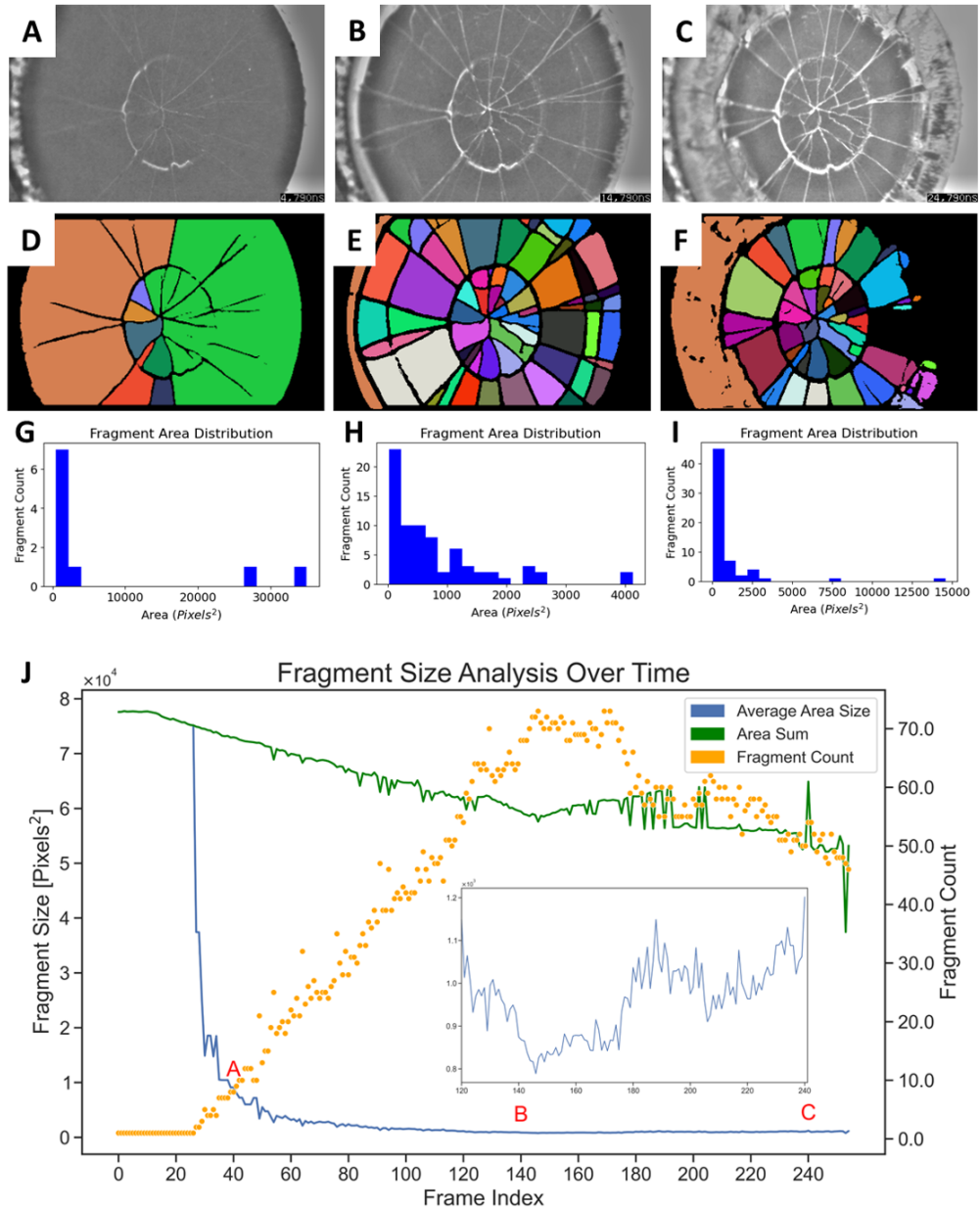

**Figure 6.** Fragment size analysis conducted over a given experiment time series. A – C display the input images and are annotated at the corresponding frame index in the graph. D – F exhibit the fragments retrieved from the prediction model. The respective fragment size distribution is included (G-I). Finally, (J) shows the evolution in fragment size, total fragment area and fragment count that corresponds to the image frame index.

| Type | Model Name | Mean IoU Score | Mean Recall | Mean Dice Coefficient | Mean F1 Score | Mean Precision |
|---|---|---|---|---|---|---|
| Binary | U-Net_1_0.0001 | 0.8034 | 0.8678 | 0.8910 | 0.8910 | 0.9154 |
| Binary | Hard Ensemble | 0.7775 | 0.8667 | 0.8656 | 0.8656 | 0.8668 |
| Binary | Soft Ensemble | 0.7826 | 0.8670 | 0.8686 | 0.8686 | 0.8726 |
| Binary | BMA Ensemble | 0.7990 | 0.8789 | 0.8781 | 0.8781 | 0.8797 |
| 3 Classes | U-Net_1_0.0001 | 0.8416 | 0.9083 | 0.9119 | 0.9119 | 0.9156 |
| 3 Classes | Hard Ensemble | 0.8539 | 0.9106 | 0.9120 | 0.9120 | 0.9144 |
| 3 Classes | Soft Ensemble | 0.8145 | 0.8822 | 0.8892 | 0.8892 | 0.8983 |
| 3 Classes | BMA Ensemble | 0.8279 | 0.8915 | 0.8973 | 0.8973 | 0.9049 |
| 5 Classes | U-Net_1_0.0001 | 0.8416 | 0.9083 | 0.9119 | 0.9119 | 0.9156 |
| 5 Classes | Hard Ensemble | 0.7363 | 0.8309 | 0.8233 | 0.8233 | 0.8237 |
| 5 Classes | Soft Ensemble | 0.7363 | 0.8309 | 0.8233 | 0.8233 | 0.8237 |
| 5 Classes | BMA Ensemble | 0.7363 | 0.8309 | 0.8233 | 0.8233 | 0.8237 |
| All | BMA Ensemble | 0.7225 | 0.9454 | 0.8367 | 0.8367 | 0.7517 |

**Table 1.** Performance Metrics for Different Models